\documentclass[pra,reprint,bibnotes,floatfix]{revtex4-2}  

\usepackage[T1]{fontenc}                        
\usepackage[australian]{babel}
\usepackage[a4paper,centering,hmargin=1.75cm,vmargin=2cm]{geometry} 
\usepackage{amsmath,amssymb,graphicx,bm,microtype} 
\usepackage[dvipsnames]{xcolor}
\usepackage[colorlinks,allcolors=blue!50!black]{hyperref}
\usepackage[all]{hypcap} 
\usepackage{braket,cleveref,enumitem,titlesec,siunitx,newfloat}

\titleformat{\paragraph}[runin]{\bfseries}{}{0pt}{}[.]
\DeclareFloatingEnvironment[
  fileext = loa,
  listname = {List of algorithms},
  name = Algorithm
]{algorithm}
\newcommand{\abs}[1]{\lvert #1 \rvert}
\renewcommand{\vec}[1]{\bm{\mathrm{#1}}}
\let\Re\relax \DeclareMathOperator{\Re}{Re}

\begin{document}

\title{Mechanism of delocalisation-enhanced exciton transport in disordered organic semiconductors}

\author{Daniel Balzer}
\affiliation{School of Chemistry, University of Sydney, NSW 2006, Australia}

\author{Ivan Kassal}
\email[Email: ]{ivan.kassal@sydney.edu.au}
\affiliation{School of Chemistry, University of Sydney, NSW 2006, Australia}

\begin{abstract}
Large exciton diffusion lengths generally improve the performance of organic semiconductor devices, since they enable energy to be transported farther during the exciton lifetime. However, the physics of exciton motion in disordered organic materials is not fully understood, and modelling the transport of quantum-mechanically delocalised excitons in disordered organic semiconductors is a computational challenge. Here, we describe delocalised kinetic Monte Carlo (dKMC), the first model of three-dimensional exciton transport in organic semiconductors that includes delocalisation, disorder, and polaron formation. We find that delocalisation can dramatically increase exciton transport; for example, delocalisation across less than two molecules in each direction can increase the exciton diffusion coefficient by over an order of magnitude. The mechanism for the enhancement is twofold: delocalisation enables excitons both to hop more frequently and further in each hop. We also quantify the effect of transient delocalisation (short-lived periods where excitons become highly delocalised), and show it depends strongly on the disorder and the transition dipole moments.
\begin{center}
\includegraphics[width=6cm]{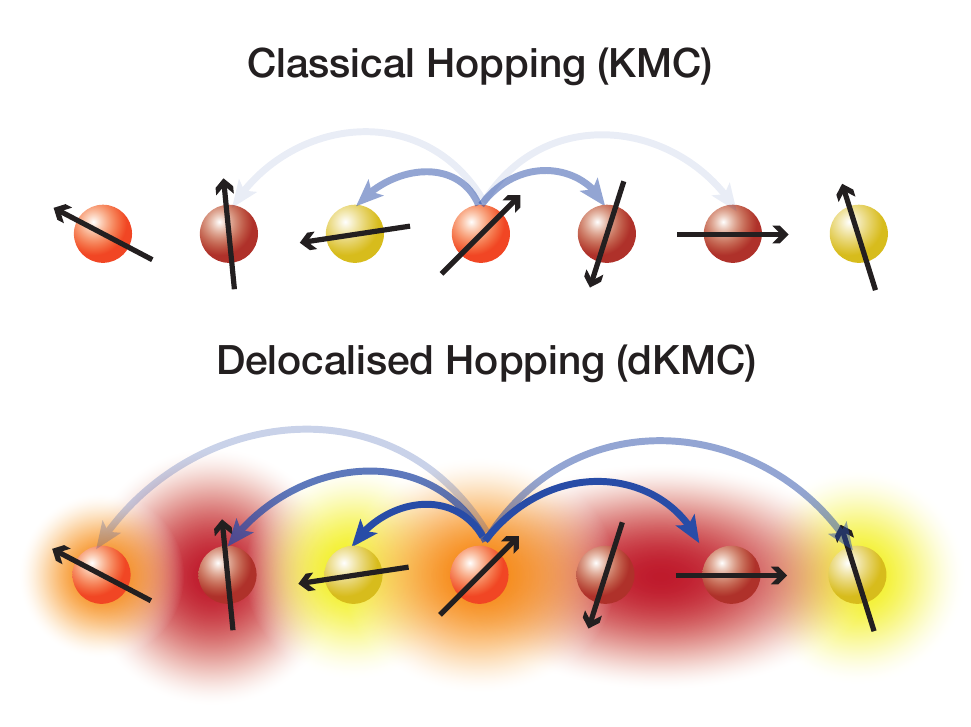}
\end{center}
\end{abstract}

\maketitle

Efficient energy transport---in the form of excitons---is essential to the performance of organic semiconductor devices, including solar cells, light emitting diodes, and flexible electronics~\cite{Bredas2004,Menke2013,Mikhnenko2015,kohler2015textbook,Bjorgaard2015,Dimitriev2022}. 
However, continued development of materials with large exciton diffusion lengths is limited by theoretical and computational models of exciton transport that lack important features. In particular, excitons are often assumed to be localised onto individual molecules and to hop between them via F\"{o}rster resonant energy transfer (FRET)~\cite{Forster1948,MayKuhn,Athanasopoulos2009,Stehr2014,kohler2015textbook,Hume2021}. However, the assumption of localised excitons often fails, leading to underestimates of how far excitons can travel~\cite{Hume2021}. Instead, the movement of excitons often falls into the theoretically awkward intermediate regime between completely localised excitons (described using FRET) and completely delocalised ones (described using band transport).

\begin{figure*}
    \centering
    \includegraphics[width=\textwidth]{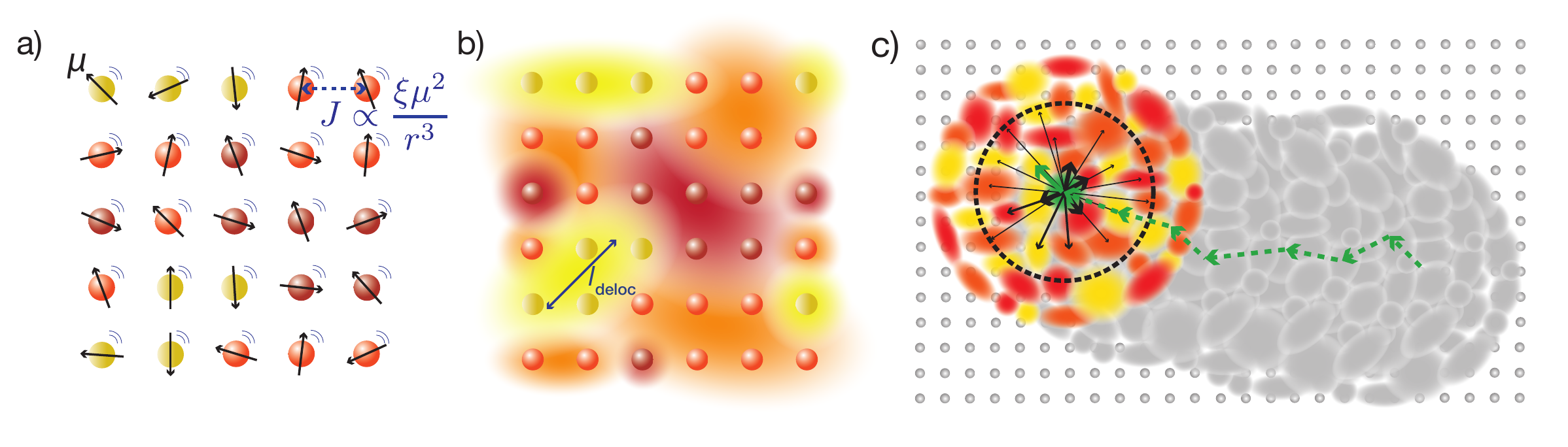}
    \caption{{\bf{dKMC model of exciton transport.}} 
    \textbf{a)} Exciton transport is modelled on a regular lattice of sites (spheres) with randomly oriented transition dipole moments $\vec{\mu}$ and disordered energies (different colours). Sites are coupled to each other with dipole-dipole coupling $J$, and to an environment (motion lines). 
    \textbf{b)} Diagonalising the polaron-transformed Hamiltonian produces the partially delocalised excitonic states (clouds; only some states shown), whose size depends on the strength of the disorder, the excitonic couplings, and the coupling to the environment. 
    \textbf{c)} dKMC propagates the dynamics through the excitonic states by tracking, and averaging over, individual trajectories (green line) through large systems. At each step, polaron states are only calculated within a small neighbourhood, and the destination state is chosen probabilistically from the outgoing rates (arrows), which are only calculated for states within a cutoff distance (black dotted line).}
    \label{fig:model}
\end{figure*}

Recent studies have significantly improved the modelling of partially delocalised excitons in the intermediate regime, showing that delocalisation improves exciton transport. These studies have ranged from detailed atomistic approaches using MCTDH~\cite{Binder2013,Wahl2014,Binder2020,Popp2021} to approaches that balance accuracy and performance to extend the simulations to larger length or time scales, including quantum master equations~\cite{Jankovic2015,Arago2016,Shi2018,Lee2019,Varvelo2021,Campaioli2021} and surface hopping~\cite{Giannini2022,Sneyd2021,Prodhan2021,Kranz2016}.
It has been proposed that delocalisation enhancements of exciton transport are caused by short periods of large delocalisation, dubbed transient delocalisation~\cite{Sneyd2021,Giannini2022,Sneyd2022}. 
In particular, Sneyd et al.\ hypothesised that large diffusion coefficients in 1D P3HT nanofiber films were a consequence of transient delocalisation, after observing individual calculated exciton trajectories in which otherwise localised excitons occasionally moved a large distance through brief transitions to highly delocalised states~\cite{Sneyd2021}.
In the context of exciton transport in organic crystals, Giannini et al.\ showed that ignoring delocalisation events reduced the diffusion coefficient three-fold~\cite{Giannini2022}. 

However, the basic mechanism of delocalisation-enhanced exciton transport remains incompletely understood, largely because computational complexity of existing techniques limits simulations to individual materials, low dimensions, short times, short length scales, or few trajectories. Conclusively establishing that diffusion enhancements are caused by delocalisation requires a method that can go beyond studying individual materials and can predict trends across wide parameter ranges, while reproducing localised hopping in the correct limits. Similarly, determining whether delocalisation enhancements are caused by large enhancements to a few events (as suggested by the hypothesis of transient delocalisation) or smaller enhancements to many events requires a way to quantify transient delocalisation. And to understand the role of delocalisation in organic devices requires the ability to model delocalised exciton motion in mesoscopic three-dimensional systems over realistic transport timescales.

Here, we solve these problems by developing delocalised kinetic Monte Carlo (dKMC), the first model of three-dimensional exciton transport that includes the essential ingredients of disorder, delocalisation, and polaron formation. Our algorithm is based on our dKMC for charge transport~\cite{Balzer2020,Balzer2022}; although the equations of motion are similar, exciton dynamics is significantly different than that of charges because of the long-range nature of excitonic couplings.
The numerical performance of dKMC allows us to scan wide parameter ranges to establish that delocalisation improves exciton motion on long time and length scales, and in the three dimensions inaccessible to some previous techniques. 
We show that the delocalisation enhancement is a consequence of both increased hopping distances and frequencies, and not just one factor alone. 
Lastly, we develop a general method to quantify the contributions of transient delocalisation events, showing that the impact of transient delocalisation depends strongly on the energetic disorder and the molecular transition dipole moments.

We model the transport of excitons on a regular, $d$-dimensional lattice of $N^d$ sites  (\cref{fig:model}a). The energies $E_n$ of the sites are disordered, chosen randomly from a Gaussian distribution, $g(E)=\exp\left(-(E-E_0)^2/2\sigma^2\right)/\sqrt{2\pi\sigma^2},$ whose standard deviation $\sigma$ is the excitonic disorder~\cite{Bassler1993}. Each site is also assigned a transition dipole moment $\vec{\mu}_n$, with constant magnitude $\mu$ but random orientation. The sites are coupled to each other by the dipole-dipole interaction, $J_{mn}=\xi_{mn}\mu^2/4\pi \epsilon_0 \abs{\vec{R}_{mn}}^3$, where $\vec{R}_{mn}$ is the distance vector between sites $m$ and $n$ and $\xi_{mn}$ is the orientation factor, $\xi_{mn}=\hat{\vec{\mu}}_m\cdot\hat{\vec{\mu}}_n - 3( \hat{\vec{R}}_{mn}\cdot\hat{\vec{\mu}}_m)(\hat{\vec{R}}_{mn}\cdot\hat{\vec{\mu}}_n)$, where hats indicate corresponding unit vectors. The Hamiltonian describing the system is then given by 
\begin{equation}
    H_\mathrm{S} = \sum_n E_n \ket{n}\bra{n} + \sum_{m \neq n} J_{mn} \ket{m}\bra{n}.
\end{equation}

The environment is treated as an independent bath of harmonic oscillators on every site~\cite{kohler2015textbook,MayKuhn}, 
\begin{equation}
    H_\mathrm{B}=\sum_{n,k}\omega_{nk} b^\dag_{nk}b_{nk},
\end{equation}
where $\omega_{nk}$ is the frequency of mode $k$ on site $n$, with creation and annihilation operators $b^\dag_{nk}$ and $b_{nk}$.
The system-bath interaction is described with a linear coupling of each site to its bath modes,
\begin{equation}
    H_\mathrm{SB} = \sum_{n,k} g_{nk}\ket{n}\bra{n}(b^\dag_{nk} + b_{nk}).
\end{equation}

We account for the formation of (excitonic) polarons, quasi-particles containing the exciton and the associated distortion to the bath~\cite{Frolich1954,Holstein1959}. Polaron formation is included in the model by applying the polaron transformation~\cite{Grover1971}, 
$e^S = \exp\big(\sum_{n,k}\frac{g_{nk}}{\omega_{nk}}\ket{n}\bra{n}(b^\dag_{nk}-b_{nk})\big)$ to the entire Hamiltonian, giving
$\tilde{H}_\mathrm{tot} = e^S H_\mathrm{tot}e^{-S} = \tilde{H}_\mathrm{S} + \tilde{H}_\mathrm{B} + \tilde{H}_\mathrm{SB}$. The transformation displaces the bath modes, giving the transformed system Hamiltonian
\begin{equation}
\label{eq:pH_S}
\tilde{H}_\mathrm{S} = \sum_n \tilde{E}_n \ket{n}\bra{n} + \sum_{m \neq n} J_{mn}\kappa_{mn} \ket{m}\bra{n},
\end{equation} 
where $\tilde{E}_n=E_n-\sum_k \abs{g_{nk}}^2/\omega_k$ and the excitonic couplings are renormalised by
\begin{equation}
\label{eq:kappa}
    \kappa_{mn} = e^{-\frac{1}{2}\sum_k\left(\frac{g^2_{mk}}{\omega^2_{mk}}\coth{\left(\frac{\omega_{mk}}{2k_{\mathrm{B}}T}\right)}+\frac{g^2_{nk}}{\omega^2_{nk}}\coth{\left(\frac{\omega_{nk}}{2k_{\mathrm{B}}T}\right)}\right)},
\end{equation}
where $T$ is the temperature.
The bath Hamiltonian is unaffected, $\tilde{H}_\mathrm{B} = H_\mathrm{B}$, and the system-bath Hamiltonian becomes
$\tilde{H}_\mathrm{SB} = \sum_{n\ne m} J_{mn}\ket{m}\bra{n}V_{mn}$,
where 
$V_{mn} = \exp\big(\sum_k\frac{g_{mk}}{\omega_{mk}}(b^\dag_{mk}-b_{mk}) -\sum_k\frac{g_{nk}}{\omega_{nk}}(b^\dag_{nk}-b_{nk})\big) - \kappa_{mn}$.

For simplicity, we assume the same system-bath interaction on all sites, $g_{nk}=g_k$, with spectral density $J(\omega)=\sum_kg_k^2\delta(\omega-\omega_k)$~\cite{MayKuhn}. Doing so simplifies the renormalisation factor to
$\kappa_{mn} = \kappa = \exp\big(-\int_0^\infty d\omega\frac{J(\omega)}{\omega^2}\coth{\left(\omega/2k_{\mathrm{B}}T\right)}\big)$.
We use a super-Ohmic spectral density $J(\omega) = \frac{\lambda}{2} (\omega/\omega_c)^3 \exp(-\omega/\omega_c)$~\cite{Pollock2013,Jang2011,Jang2002,Wilner2015},
with a reorganisation energy of $\lambda=\SI{150}{meV}$ of each molecule (which is within the range of typical values found in organic semiconductors~\cite{Hume2021,Campaioli2021,Kranz2016,Giannini2022}) and a cutoff frequency $\omega_{c}=\SI{62}{meV}$~\cite{Lee2012}.

\begin{figure*}
    \centering
    \includegraphics[width=0.9\textwidth]{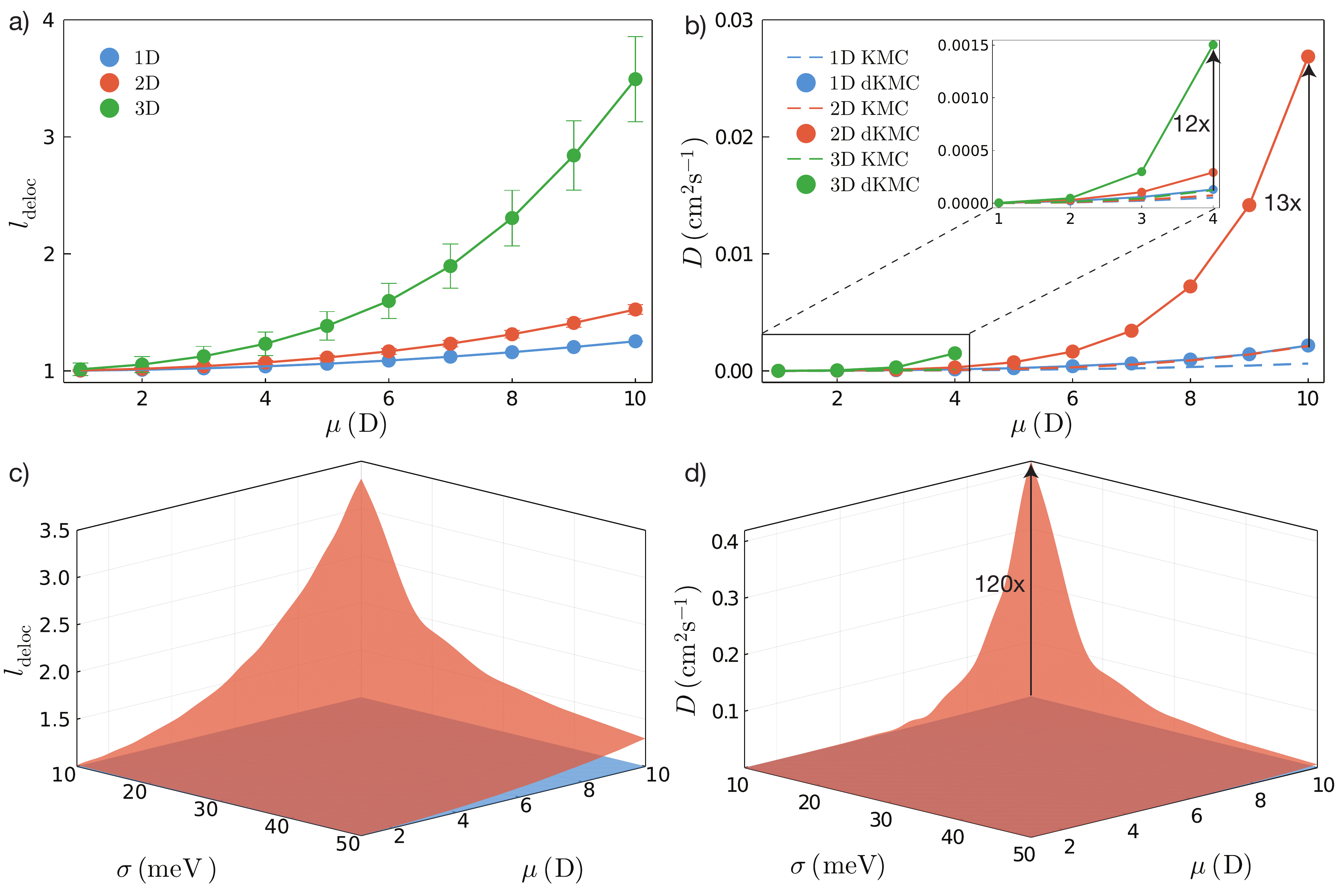}
    \caption{{\bf{Delocalisation enhances exciton transport.}} 
    \textbf{a)} Delocalisation length $l_\mathrm{deloc}$ of excitonic states increases with the transition dipole moment $\mu$ in all dimensions (shown for $\sigma=\SI{30}{meV}$).
    \textbf{b)} Exciton diffusion coefficient $D$ as a function of $\mu$, predicted without (KMC) and with (dKMC) delocalisation, for $\sigma=\SI{30}{meV}$. In all dimensions, the dKMC diffusion coefficient is larger and increases faster with $\mu$ compared to that predicted by KMC. This delocalisation enhancement is greater in higher dimensions. The error bars are the standard errors of the mean.
    \textbf{c)} The delocalisation length $l_\mathrm{deloc}$ and \textbf{d)} diffusion coefficient $D$ predicted by KMC (blue) and dKMC (orange), for varying $\sigma$ and $\mu$ in two dimensions. Both the extent of delocalisation ($l_\mathrm{deloc}$) and the delocalisation enhancement of $D$ increase with both increasing $\mu$ and decreasing $\sigma$.
    }
    \label{fig:delocalisation_enhancement}
\end{figure*}

After the polaron transformation, we diagonalise $\tilde{H}_S$ to find the polaronic states $\nu$ (\cref{fig:model}b). Because $\kappa<1$, the polaron transformation reduces the excitonic couplings, meaning that polaronic states are smaller than those of the bare excitons~\cite{Rice2018}, simplifying the calculation. The polaron transformation also absorbs most of the system-bath interaction into the polaron itself, after which the reduced system-bath coupling is treated as a perturbation to second order~\cite{Grover1971,Jang2008,Nazir2009,Jang2009,Jang2011,McCutcheon2011,Kolli2011,McCutcheon2011_2,McCutcheon2011_3,Pollock2013,Lee2015,Xu2016}. As detailed previously~\cite{Lee2015,Balzer2020}, the result of the perturbative treatment is the secular polaron-transformed Redfield equation (sPTRE)~\cite{Lee2015},
\begin{equation}
\label{eq:sPTRE}
\frac{d\rho_{\nu\nu}(t)}{dt} = \sum_{\nu'}R_{\nu\nu'}\rho_{\nu'\nu'}(t),
\end{equation}
a master equation for the polaron-state populations $\rho_{\nu\nu}(t)$. The Redfield transition rates
$R_{\nu\nu'} = 2\Re\Gamma_{\nu'\nu,\nu\nu'} - \delta_{\nu\nu'}\sum_\kappa2\Re\Gamma_{\nu\kappa,\kappa\nu}$
describe the bath-induced relaxation in terms of the damping rates
\begin{multline}
\label{eq:Gamma}
\Gamma_{\mu\nu,\mu'\nu'} = \sum_{m,n,m',n'} J_{mn} J_{m'n'} 
\\ \braket{\mu|m}\braket{n|\nu}\braket{\mu'|m'}\braket{n'|\nu'} K_{mn,m'n'}(\omega_{\nu'\mu'}),
\end{multline}
where
$K_{mn,m'n'}(\omega) = \int_0^\infty e^{i\omega \tau}\braket{\tilde{\hat{V}}_{mn}(\tau)\tilde{\hat{V}}_{m'n'}(0)}_\mathrm{B}d\tau$
is the half-Fourier transform of the bath correlation function
$\braket{\tilde{\hat{V}}_{mn}(\tau)\tilde{\hat{V}}_{m'n'}(0)}_{\mathrm{B}} = \kappa^2(e^{\lambda_{mn,m'n'}\phi(\tau)} - 1)$,
where $\lambda_{mn,m'n'} = \delta_{mm'} - \delta_{mn'} + \delta_{nn'} - \delta_{nm'}$ and 
$\phi(\tau) = \int_0^\infty d\omega\frac{J(\omega)}{\omega^2}\left(\cos(\omega \tau)\coth(\beta\omega/2) - i \sin(\omega \tau)\right)$~\cite{Jang2011}.

\begin{figure*}
    \centering
    \includegraphics[width=0.9
    \textwidth]{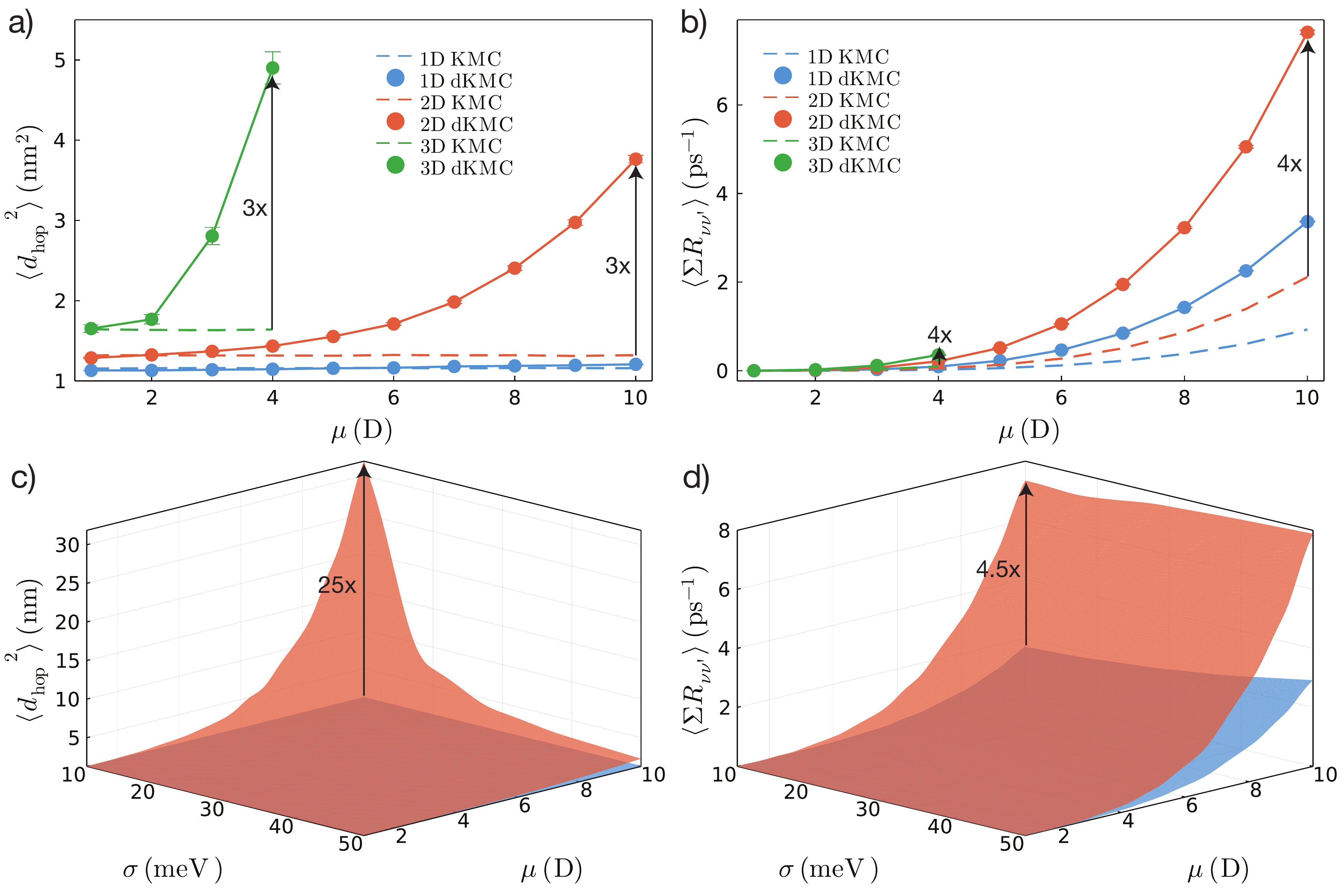}
    \caption{{\bf{Mechanism of delocalisation-enhanced exciton transport.}} 
    \textbf{a)}~The mean squared hopping distance $\langle d_\mathrm{hop}^2\rangle$ and \textbf{b)}~mean outgoing rate sum $\langle \Sigma_{\nu'} R_{\nu\nu'} \rangle$ as functions of the transition dipole moment $\mu$ for dKMC and KMC in each dimension. As delocalisation increases with increasing $\mu$, the dKMC values of both $\langle d_\mathrm{hop}^2\rangle$ and $\langle \Sigma_{\nu'} R_{\nu\nu'} \rangle$ grow faster than their KMC (localised-hopping) versions. When the enhancements to $\langle d_\mathrm{hop}^2\rangle$ and $\langle \Sigma_{\nu'} R_{\nu\nu'} \rangle$ are multiplied, they account for the total delocalisation enhancement to $D$ seen in \cref{fig:delocalisation_enhancement}b. Similarly,
    \textbf{c)}~$\langle d_\mathrm{hop}^2\rangle$ and \textbf{d)}~$\langle \Sigma_{\nu'} R_{\nu\nu'} \rangle$ as functions of both $\mu$ and the disorder $\sigma$ for dKMC (orange) and KMC (blue) in two dimensions. As in a--b, multiplying the enhancements to $\langle d_\mathrm{hop}^2\rangle$ and $\langle \Sigma_{\nu'} R_{\nu\nu'} \rangle$ accounts for the delocalisation enhancement to $D$ in \cref{fig:delocalisation_enhancement}d.}
    \label{fig:enhancement_mechanism}
\end{figure*}

Solving \cref{eq:sPTRE} to calculate the dynamics of all the excitons is only computationally tractable for small systems. Instead, we use dKMC~\cite{Balzer2020,Balzer2022}, which stochastically unravels the master equation onto kinetic Monte Carlo and improves numerical performance using distance-based cutoffs enabled by the limited polaron sizes. Here, we summarise the algorithm, given in full in \cref{sec:dKMC_appendix}. 
First, we select a random disordered landscape of $N^d$ sites from the distributions described above; then, and throughout the simulation, we diagonalise subsets of this landscape to only find polaron states close to the current location of the exciton (\cref{fig:model}c).
Then, Redfield rates $R_{\nu\nu'}$ for hopping from the current state $\nu$ are only calculated for destination states $\nu'$ that lie within a certain cutoff radius. 
In calculating each of these rates, we truncate the sum in \cref{eq:Gamma} to only include sites contributing the most to the populations of each state, i.e., the fewest sites $m$ such that $\sum_m\abs{\braket{\mu|m}}^2$ exceeds a population cutoff.
Both of these cutoffs are adjusted to obtain a target accuracy. 
The destination state is chosen probabilistically among the possible targets, in proportion to the corresponding hopping rates, as in standard kinetic Monte Carlo. 
The procedure is repeated until a chosen final time $t_\mathrm{end}$, giving an individual exciton trajectory. 
The simulation is then repeated for many trajectories on many realisations of disorder to obtain sufficient statistics and allow the exciton diffusion coefficient to be calculated as
\begin{equation}
    \label{eq:D}
    D=\left.\frac{d}{dt}\left(\frac{\overline{\langle r^2(t)\rangle}}{2d}\right)\right|_{t=t_\mathrm{end}}
\end{equation}
where $\overline{\langle r^2(t)\rangle}$ is the mean-squared exciton displacement, averaged over trajectories (angle brackets) and realisations of disorder (overline).

All approximations in dKMC are conservative, i.e., they underestimate the extent and effect of delocalisation, as detailed in \cref{sec:appendix_parameter_choice}. In particular, distance-based cutoffs required by dKMC lead to an underestimation of delocalisation effects, as does a finite simulation time scale $t_\mathrm{end}$. In this work, we use $t_\mathrm{end}=\SI{100}{ps}$, a typical excitonic transit time for typical length scales in organic semiconductors. Although some excitons move on longer time scales, increasing $t_\mathrm{end}$ would also increase delocalisation enhancements. 

To determine the effect of delocalisation on exciton transport, we vary two key parameters, the transition dipole moment $\mu$ and the excitonic disorder $\sigma$. \Cref{fig:delocalisation_enhancement}a shows that increasing $\mu$ increases the exciton delocalisation, especially in higher dimensions. We quantify the delocalisation using the delocalisation length
\begin{equation}
    \label{eq:l_deloc}
    l_\mathrm{deloc} = \overline{\mathrm{IPR}_\nu}^{1/d},
\end{equation}
where $\overline{\mathrm{IPR}_\nu}$ is the average inverse participation ratio of the polaron states, 
\begin{equation}
    \label{eq:IPR}
    \mathrm{IPR}_\nu = \Big(\sum_n \abs{\braket{n|\nu}}^4\Big)^{-1}.
\end{equation}
The IPR roughly corresponds to the number of sites an excitonic state extends over. Therefore, $l_\mathrm{deloc}$ measures to extent of an excitonic state in each direction, enabling comparisons between different dimensions. The considerably larger $l_\mathrm{deloc}$ in 3D indicates the importance of fully three-dimensional simulations.

Delocalisation significantly increases exciton diffusion, especially in higher dimensions (\cref{fig:delocalisation_enhancement}b). In each dimension, the diffusion coefficients predicted by KMC and dKMC agree at low $\mu$, where the electronic states are localised. As $\mu$ increases, $D_\mathrm{dKMC}$ becomes larger than $D_\mathrm{KMC}$, demonstrating that delocalisation enhances exciton transport. For $\mu=\SI{10}{D}$ in two dimensions, delocalisation across less than two molecules in each direction ($l_\mathrm{deloc}=1.5$) gives a delocalisation enhancement of $D_\mathrm{dKMC}/D_\mathrm{KMC}=13$. Furthermore, the delocalisation enhancement is greater in higher dimensions if all parameters are held fixed.
In three dimensions, despite the large computational savings, dKMC is limited to small $\mu$ because the excitonic states become too large to be contained within computationally tractable boxes.

\begin{figure*}
    \centering
    \includegraphics[width=\textwidth]{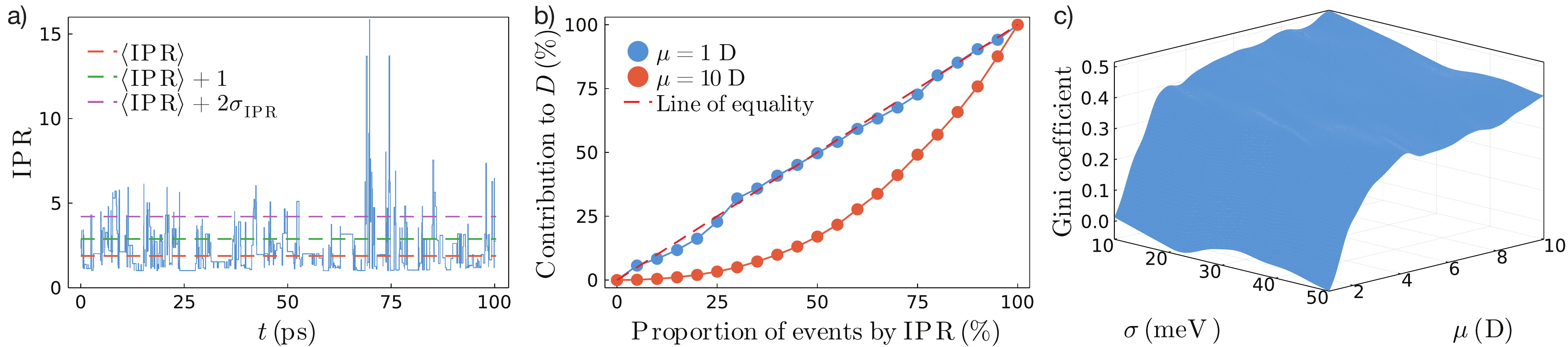}
    \caption{{\bf{Quantifying transient delocalisation.}} \textbf{a)} IPR of an exciton over a single trajectory in 2D with $\mu=\SI{10}{D}$. 
    Most of the time, the exciton's IPR is near the average $\langle \mathrm{IPR}\rangle$, but occasionally it briefly delocalises a large amount, known as transient delocalisation. 
    \textbf{b)} Lorenz curves showing the cumulative contributions to $D$ of the bottom $x\%$ of hopping events, ranked by IPR. The line of equality represents exciton transport where every hopping event contributes equally to $D$. The Gini coefficient quantifies how disproportionate the contributions from highly delocalised events are; it is twice the area between the Lorenz curve and the line of equality. In the more localised example ($\mu=\SI{1}{D}$), a low Gini coefficient $G=0.01$ implies insignificant transient delocalisation. In the more delocalised example ($\mu=\SI{10}{D}$), the higher value $G=0.43$ demonstrates a greater importance for transient delocalisation events; for instance, 25\% of the most delocalised hopping events account for 51\% of $D$. 
    \textbf{c)} Gini coefficients for a range of $\mu$ and $\sigma$ in 2D, showing that the importance of transient delocalisation depends strongly on the parameters.}
    \label{fig:transient_delocalisation}
\end{figure*}

Our conclusions about the importance of delocalisation are general, holding at all typical values of excitonic disorder. In particular, at any $\sigma$, increasing $\mu$ increases the IPR, the diffusion coefficient $D$, and the enhancement $D_\mathrm{dKMC}/D_\mathrm{KMC}$ (shown for two dimensions in \cref{fig:delocalisation_enhancement}c--d). These parameter scans also show the deleterious effect of disorder on exciton diffusion; increasing $\sigma$ reduces both IPR and $D$ at constant $\mu$. 

The mechanism of delocalisation-enhanced exciton transport is twofold: excitons both hop further in each hop and they hop more frequently, two contributions that are distinguished in \cref{fig:enhancement_mechanism}. The first mechanism is that delocalised excitons hop further in each hop, on average, than localised ones. \Cref{fig:enhancement_mechanism}a shows that the mean squared hopping distance $\langle d_\mathrm{hop}^2 \rangle$ grows as a function of $\mu$ in both KMC and dKMC, but much more rapidly for the latter. The parameter scan in \cref{fig:enhancement_mechanism}c repeats the same calculation at various values of $\sigma$, showing that $\langle d_\mathrm{hop}^2 \rangle$ increases with both increasing $\mu$ and decreasing $\sigma$. As $\mu$ increases or $\sigma$ decreases (or both), the excitons become more delocalised, increasing their coupling to exciton states that are further away, thus enabling them to hop further in one hop. Furthermore, the longer-distance couplings provide more possible hopping destinations, increasing the likelihood of finding an energetically favourable (and thus faster) transition. 
The second mechanism of delocalisation-enhanced transport is that the rate of hopping between delocalised excitons is greater on average, reducing the time between transitions. \Cref{fig:enhancement_mechanism}b shows the average sum of transition rates for hops leaving a particular state, $\langle \Sigma_{\nu'} R_{\nu\nu'} \rangle$, as a function of $\mu$. As $\mu$ increases, $\langle \Sigma_{\nu'} R_{\nu\nu'} \rangle$ increases in both KMC and dKMC, but considerably faster in dKMC. The calculation is repeated in \cref{fig:enhancement_mechanism}d for $\langle \Sigma_{\nu'} R_{\nu\nu'} \rangle$, but as a function of both $\mu$ and $\sigma$. 

Both mechanisms are necessary to explain the large delocalisation enhancements seen in \cref{fig:delocalisation_enhancement}b,d. For example, for the parameter values in \cref{fig:enhancement_mechanism}a--b, the enhancement to $\langle d_\mathrm{hop}^2 \rangle$ is $3\times$, while that to $\langle \Sigma_{\nu'} R_{\nu\nu'} \rangle$ is $4\times$, neither of which is sufficent to explain the $12\times$ or $13\times$ enhancement to $D$ in \cref{fig:delocalisation_enhancement}b. However, by dimensional analysis, the enhancement to $D$ should be proportional to the product of the enhancements to $\langle d_\mathrm{hop}^2 \rangle$ and $\langle \Sigma_{\nu'} R_{\nu\nu'} \rangle$. Indeed, the product of the $3\times$ and $4\times$ enhancements accounts for the overall 12--$13\times$ enhancement in $D$. Similarly, over the greater parameter range in \cref{fig:enhancement_mechanism}c--d, multiplying the $25\times$ enhancement to $\langle d_\mathrm{hop}^2 \rangle$ with the $4.5\times$ enhancement to $\langle \Sigma_{\nu'} R_{\nu\nu'} \rangle$ explains the overall $120\times$ enhancement to $D$ seen in \cref{fig:delocalisation_enhancement}d.

Although the average enhancements can be explained using the mechanisms above, averages do not tell the whole story, and to understand transport, we also need to look at distributions. An exciton's delocalisation is not fixed, but can fluctuate rapidly and by a large amount from the mean $\langle \mathrm{IPR}\rangle$ (\cref{fig:transient_delocalisation}a). Transient delocalisation is the hypothesis that these fluctuations are important, i.e., that a few hops involving highly delocalised states contribute disproportionately to the delocalisation enhancement to $D$, whereas the alternative would be that the delocalisation enhancement was caused by smaller improvements to many (or all) of the hops. 

Distinguishing these two possibilities requires a way to measure the inequality of distributions. To do so, we use the Lorenz curve, a plot of the cumulative distribution function commonly used in economics to quantify the inequality of wealth distributions~\cite{Lorenz1905,Gini1912}. In studying wealth distributions, the Lorenz curve shows the fraction of the wealth owned by the bottom $x\%$ of the population, i.e., the cumulative proportion of the total wealth held by a cumulative proportion of the total population (ranked by wealth). If every person had an equal amount of wealth, the Lorenz curve would be a straight line known as the line of equality. Departures from equal distributions are measured by the Gini coefficient $G$, which is twice the area between the line of equality and the Lorenz curve. A fully equal wealth distribution has a Gini coefficient of 0, and the wealth inequality grows as the Gini coefficient increases, up to its maximum value of 1. 

To quantify the inequality of distributions of exciton hops based on the extent of delocalisation, we plot the cumulative contribution to the diffusion constant $D$ of the cumulative proportion of hopping events, ranked by IPR (\cref{fig:transient_delocalisation}b). To construct this Lorenz curve, we assign to each hop the greater of the IPRs of the donor and acceptor states; then, we rank all the hops in all the trajectories based on this IPR. To calculate the contribution of the bottom $x\%$ of hops, we construct new trajectories where the top $(100-x)\%$ of hops are removed, i.e., we connect together the displacement vectors of the retained hops. The diffusion coefficient predicted by these new trajectories is the contribution to the total $D$ that can be assigned to the bottom $x\%$ hops.
Unlike Lorenz curves for wealth inequality, our Lorenz curve may, in exceptional cases, rise above the line of equality because it is possible (although rare) for states with smaller IPR to contribute more to $D$ than larger states. Nevertheless, the Gini coefficient remains a useful measure of the disproportionate influence of transient delocalisation effects.

\Cref{fig:transient_delocalisation}b--c shows that the Gini coefficent is smaller for relatively localised systems and larger for delocalised ones, whether the delocalisation is caused by large $\mu$ or small $\sigma$ (or both). Therefore, the importance of transient delocalisation depends strongly on the parameter regime: in disordered systems with weak couplings it can be negligible, and it only becomes significant in organic semiconductors that are relatively ordered and have strong couplings among sites. This agrees with the finding that transient delocalisation can have a large effect in organic crystals~\cite{Giannini2022}, where disorder is low and couplings are large.

Quantifying transient delocalisation with the Gini coefficient has the advantage of taking into account the entire distribution of trajectories. By contrast, initial approaches to transient delocalisation~\cite{Sneyd2021} only identified individual transient delocalisation events in particular trajectories, which cannot be guaranteed to be typical, especially in disordered materials where individual trajectories can behave very differently.
More recent work classified events as transient delocalisation or not~\cite{Giannini2022}; however, doing so requires an arbitrary cutoff and discards information contained in the full distribution. For example, IPR fluctuations to $\langle\mathrm{IPR}\rangle+1$ may be sufficiently rare to classify such events as transient delocalisation in one material~\cite{Giannini2022}, but insufficient in another material with larger fluctuations. For example, \cref{fig:transient_delocalisation}a shows a material where a cutoff of $\mathrm{IPR}=\langle\mathrm{IPR}\rangle +1$ would label as many as 24\% of events as transient delocalisation, meaning that they are no longer remarkable, rare events. A more generic definition might have a cutoff that depends on the spread of the IPRs, setting the limit of transient delocalisation at $\mathrm{IPR}=\langle\mathrm{IPR}\rangle +w\sigma_\mathrm{IPR}$, but even this requires an arbitrary choice of how many standard deviations $w$ should be considered. Because our method applies to any distribution of events, it can quantify transient delocalisation in all parameter regimes (\cref{fig:transient_delocalisation}). 

Overall, the computational performance of dKMC has enabled some of the largest simulations of delocalised exciton transport in disordered materials. Previously, the largest such simulations were 2D simulations of about 300 molecules for $\SI{1}{ps}$~\cite{Giannini2022}; by contrast, we simulated 3D systems with millions of sites for $\SI{100}{ps}$. In addition, the speed of dKMC allows predictions of general trends over large parameter ranges, which have yielded the mechanistic insights above.

This gain in computational performance requires a series of approximations that can be limiting in some cases. Many of the assumptions in dKMC come from the underlying master equation, sPTRE. While sPTRE is accurate in the parameter ranges studied here, it loses accuracy when the system is weakly coupled to a slow bath, where the exciton dynamics occurs faster than the bath relaxation, preventing the relaxed-bath assumption of the fully displaced polaron transformation~\cite{Lee2012,Chang2013,Pollock2013,Lee2015}. As an alternative, the variational polaron transformation~\cite{Silbey1984,Zimanyi2012,Pollock2013,Jang2022} would allow dKMC to be more accurately applied to a system weakly coupled to slow baths, or to an Ohmic or sub-Ohmic bath.
Similarly, sPTRE uses the secular approximation to neglect coherences between states and justify tracking only the polaron populations. The approximation is justified because coherences are unlikely to be induced in incoherent light~\cite{Jiang1991,Mancal2010,Brumer2012,Kassal2013,Brumer2018,Tomasi2019,Tomasi2020,Tomasi2021}, and, even if they were, they would be unlikely to survive on exciton-transport timescales. However, if required, dKMC could be adapted to include coherences.
A final assumption is the local (diagonal) system-bath coupling, which is usually made in disordered organic semiconductors~\cite{kohler2015textbook,MayKuhn}. However, in organic crystals, non-local (off-diagonal) system-bath couplings become important~\cite{Arago2016}. Relaxing the local-bath assumption would require a significant adjustment to the sPTRE equations of motions, since the polaron transformation relies on the ability to remove diagonal system-bath couplings.

dKMC could be applied to explain exciton-transport behaviour of specific disordered materials using multiscale simulation. dKMC input parameters---$\sigma$, $\mu$, $\lambda$, and $\omega_c$---can be calculated for specific materials using atomistic quantum-chemistry simulations, as has been done for other effective-Hamiltonian models of exciton transport~\cite{Kranz2016,Varvelo2021,Prodhan2021,Giannini2022}.

We anticipate that dKMC will also be applied to describe related processes in optoelectronic materials, including exciton recombination, exciton dissociation, and singlet fission.
Eventually, we expect that it will be possible to incorporate delocalisation into mesoscopic simulation of all optoelectronic processes relevant to organic electronics and---through rates predicted by dKMC or simplifications such as jKMC~\cite{jKMC}---into device-scale drift-diffusion models.

In conclusion, dKMC describes mesoscale 3D exciton transport in organic semiconductors, for the first time including the important ingredients of disorder, delocalisation, and polaron formation. Our simulations show that delocalisation significantly improves exciton transport over classical hopping, especially in higher dimensions. We showed that this enhancement is a combined effect of larger average hopping distances and outgoing rates, and we quantified the contribution of transient delocalisation, finding that its importance depends strongly on the nature of the material. We anticipate that these mechanistic insights will aid in the discovery of improved exciton-transport materials and that our simulation techniques can be further extended to address other open questions in organic optoelectronics.

\begin{acknowledgments}
We were supported by a Westpac Scholars Trust Future Leaders Scholarship, the Australian Research Council (DP220103584), the Australian Government Research Training Program, and by computational resources from the National Computational Infrastructure (Gadi) and the University of Sydney Informatics Hub (Artemis).
\end{acknowledgments}

\bibliography{bib}

\section*{Appendices} 

\appendix 

\setcounter{section}{0}
\renewcommand{\thesection}{A\arabic{section}}%
\setcounter{equation}{0}
\renewcommand{\theequation}{A\arabic{equation}}%
\setcounter{figure}{0}
\renewcommand{\thefigure}{A\arabic{figure}}%
\setcounter{table}{0}
\renewcommand{\thetable}{A\arabic{table}}%
\setcounter{algorithm}{0}
\renewcommand{\thealgorithm}{A\arabic{algorithm}}%

\section{\label{sec:dKMC_appendix}Details of dKMC}
\begin{figure*}
    \centering
    \includegraphics[width=\textwidth]{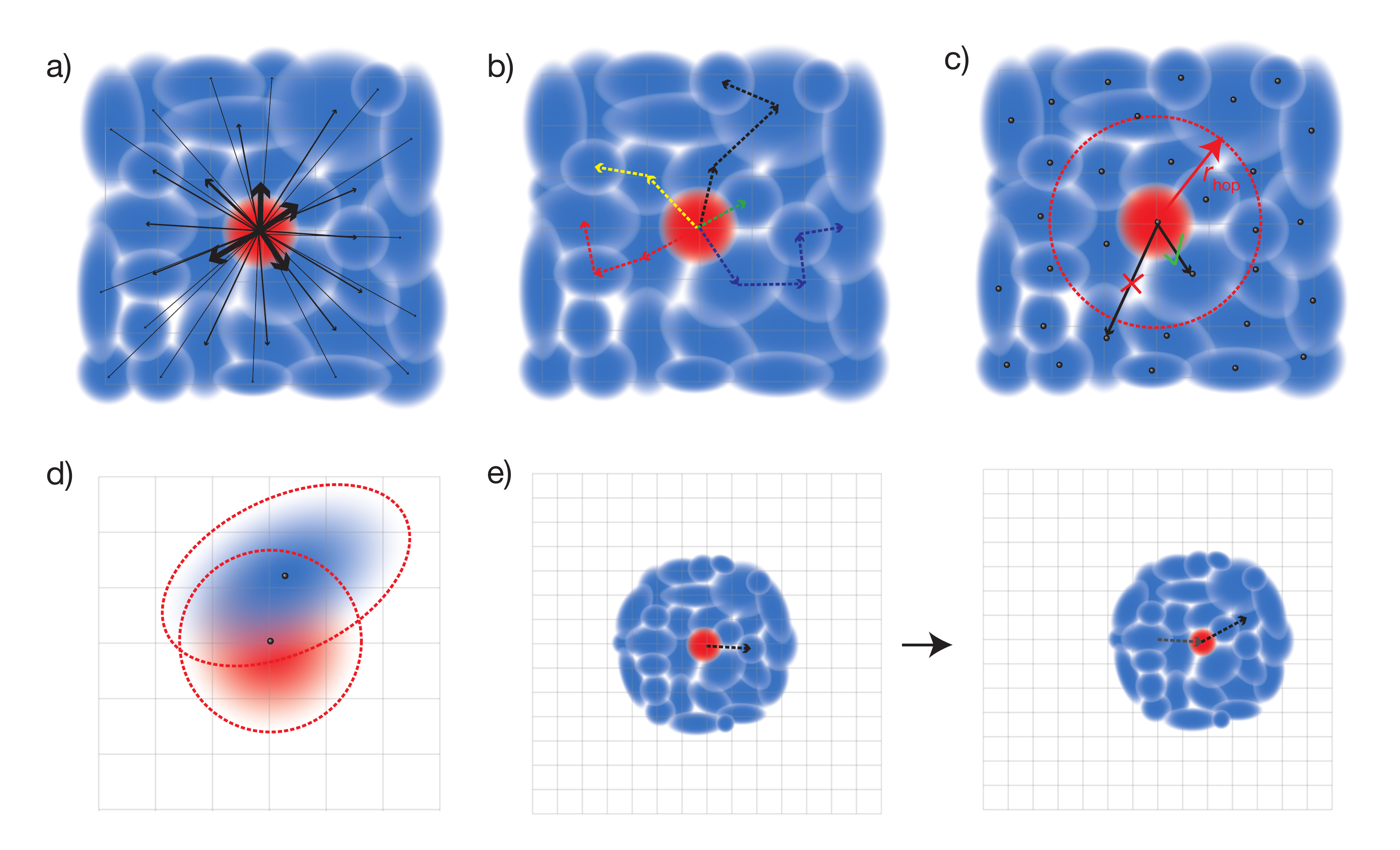}
    \caption{\textbf{Mapping sPTRE (a) onto dKMC}. \textbf{b)} KMC: many individual trajectories are formed from sequential probabilistically chosen hops, before being averaged.
    \textbf{c)} Hopping cutoff: excitons can only hop to states whose centres are within a hopping cutoff.
    \textbf{d)} Population cutoff:
    when calculating rates of transfer between excitonic states we ignore sites that do not considerably contribute to the populations of the donor and acceptor states.
    \textbf{e)} Diagonalising on the fly: Instead of calculating polaron states for the entire system, we calculate a subset surrounding the current location of the exciton after every hop.
   }
    \label{fig:approximations_excitons}
\end{figure*}

Here we detail the approximations used to map sPTRE onto dKMC (\cref{fig:approximations_excitons}) and provide the full dKMC algorithm (\cref{listing}).

\begin{algorithm*}
    \fbox{
    \begin{minipage}{0.97\textwidth}
        \setlist{nolistsep}
        \begin{minipage}[t]{0.49\textwidth}\vspace{0pt}%
            \raggedright
            Steps to be carried out for every set of microscopic parameters $N$, $d$, $\sigma$, $\mu$, $\lambda$, $\omega_c$ and $T$:
            \begin{enumerate}[leftmargin=*]
                \item (Calibrating cutoff radii) For $n_\mathrm{calib}$ realisations of disorder:
                \begin{enumerate}[leftmargin=*,label=\alph*.]
                    \item Generate an $N^d$ lattice of random energies and dipole orientations.
                    \item Set $r_\mathrm{hop}\leftarrow 0$ and $r_\mathrm{Ham}\leftarrow 0$.
                    \item While $T_{r_\mathrm{hop}-1,r_\mathrm{ove}}/T_{r_\mathrm{hop},r_\mathrm{ove}}<a_\mathrm{dKMC}$:
                    \begin{enumerate}[leftmargin=*]
                        \item Update $r_\mathrm{hop}\leftarrow r_\mathrm{hop}+1$.
                        \item While $T_{r_\mathrm{hop},r_\mathrm{Ham}-1}/T_{r_\mathrm{hop},r_\mathrm{Ham}}<a_\mathrm{dKMC}$:
                        \begin{enumerate}[leftmargin=*]
                            \item Update $r_\mathrm{ove}\leftarrow r_\mathrm{ove}+1$.
                            \item Create a polaron-transformed Hamiltonian $\tilde{H}_S$ containing all sites within a distance of $r_\mathrm{Ham}$ of the centre of the lattice and find the polaron states, their centres and their energies.
                            \item Choose polaron state $\nu$ closest to the centre of the lattice.
                            \item Create a list $L$ of all polaron states $\nu'$ such that $\abs{\vec{C}_\nu-\vec{C}_{\nu'}} < r_\mathrm{hop}$.
                            \item Calculate $R_{\nu\nu'}$ for all $\nu'\in L$, only summing in \cref{eq:Gamma} over sites containing over $a_\mathrm{dKMC}$ of the populations of each state $\nu$ and $\nu'$. 
                            \item Set $T_{r_\mathrm{hop},r_\mathrm{Ham}} \leftarrow \sum_{\nu'\in L}R_{\nu\nu'}$.
                        \end{enumerate}
                    \item Update $r_\mathrm{Ham}\leftarrow r_\mathrm{Ham}-1$.
                    \end{enumerate}
                    \item Update $r_\mathrm{hop}\leftarrow r_\mathrm{hop}-1$.
                \end{enumerate}
                \item Average $r_\mathrm{hop}$ and $r_\mathrm{Ham}$ over the $n_\mathrm{calib}$ realisations.
            \end{enumerate}
        \end{minipage}
        \hfill
        \begin{minipage}[t]{0.49\textwidth}\vspace{0pt}%
            \begin{enumerate}[leftmargin=*,start=3]
                \item (Kinetic Monte Carlo) For $n_\mathrm{iter}$ realisations of disorder:
                \begin{enumerate}[leftmargin=*, label=\alph*.]
                    \item Generate an $N^d$ lattice of random energies.
                    \item For $n_\mathrm{traj}$ trajectories:
                    \begin{enumerate}[leftmargin=*]
                        \item Create a polaron-transformed Hamiltonian $\tilde{H}_S$ containing all sites within a distance of $r_\mathrm{Ham}$ of the centre of the lattice and find the polaron states, their centres and their energies.
                        \item Set $t\leftarrow 0$ and choose initial polaron state $\nu$ closest to the centre of the lattice.
                        \item Repeat until $t>t_\mathrm{end}$:
                        \begin{enumerate}[leftmargin=*]
                            \item Create a list $L$ of all polaron states $\nu'$ such that  $\abs{\vec{C}_\nu-\vec{C}_{\nu'}} < r_\mathrm{hop}$.
                            \item Calculate $R_{\nu\nu'}$ for all $\nu'\in L$, only summing in \cref{eq:Gamma} over sites containing over $a_\mathrm{dKMC}$ of the populations of each state $\nu$ and $\nu'$.  
                            \item Set $S_{\nu'}\leftarrow\sum_{\mu=1}^{\nu'} R_{\nu\mu}$ for all $\nu'\in L$. 
                            \item Set $T\leftarrow \sum_{\nu'\in L}S_{\nu'}$.
                            \item Select the destination state by finding $\nu'$ such that $S_{\nu'-1} < uT < S_{\nu'}$, for uniform random number $u \in (0,1]$, and update $\nu\leftarrow\nu'$. 
                            \item Update $t\leftarrow t+\Delta t$, where $\Delta t = -T^{-1}\ln{v}$ for uniform random number $v \in (0,1]$.
                            \item Create a new polaron-transformed Hamiltonian $\tilde{H}_S$ containing all sites within a distance of $r_\mathrm{Ham}$ of $\vec{C}_{\nu}$, find the polaron states, their centres and their energies.
                        \end{enumerate}
                    \end{enumerate}
                \end{enumerate}
                \item Calculate $D$ using \cref{eq:D}. 
            \end{enumerate}
        \end{minipage}
    \end{minipage}
    }
    \caption{The delocalised kinetic Monte Carlo algorithm for exciton transport.}
    \label{listing}
\end{algorithm*}

In principle, the full sPTRE master equation (\cref{eq:sPTRE}) could be used to track the time evolution of the populations of all the polaron states. This is illustrated in \cref{fig:approximations_excitons}a, where the initial exciton state spreads to every other state in proportion to the Redfield rate of population transfer. The full solution to the sPTRE is
\begin{equation}
     \label{eq:sPTRE_solution}
     \rho(t)=\exp(Rt)\rho(0),
\end{equation}
where the density matrix $\rho(t)$ contains excitonic populations. The mean-squared displacement is then
$\langle r^2(t)\rangle = \mathrm{tr}(r^2\rho(t))$,
allowing $D$ to be calculated using \cref{eq:D}.

However, sPTRE is expensive and can only be applied to small systems because of three major computational tasks. First, finding the polaron states involves diagonalising the polaron-transformed system Hamiltonian (\cref{eq:pH_S}), which scales as $O(N^{3d})$. Second, calculating the full secular Redfield tensor involves calculating $O(N^{2d})$ rates between every pair of polaron states. Third, calculating each of the $O(N^{2d})$ Redfield rates requires calculating damping rates (\cref{eq:Gamma}) that include excitonic couplings and amplitudes of $O(N^{4d})$ site combinations. Therefore, the total method scales as $O(N^{3d})+O(N^{6d})$.

We apply four approximations (\cref{fig:approximations_excitons}b-e) to transform sPTRE into dKMC: mapping sPTRE onto KMC, imposing a hopping cutoff, imposing a population cutoff, and diagonalising the Hamiltonian on the fly.

First, dKMC stochastically unravels sPTRE onto KMC (\cref{fig:approximations_excitons}b). Instead of calculating the full time evolution of $\rho$, we calculate and average $n_\mathrm{traj}$ stochastic trajectories. Individual trajectories are found by probabilistically choosing the next destination and waiting time based on the Redfield rates, as in ordinary KMC. Therefore, instead of calculating Redfield rates between every pair of states, we only calculate the outgoing rates from the current state before every one of the $n_\mathrm{hop}$ hops. This reduces the number of rates that need to be calculated to $O(N^dn_\mathrm{traj}n_\mathrm{hop})$.

Second, we impose a hopping cutoff $r_\mathrm{hop}$ (\cref{fig:approximations_excitons}c). Because excitonic couplings drop off as $r^{-3}$, the outgoing rates are dominated by short hops, allowing us to neglect the long ones. Therefore, we do not calculate every outgoing rate, but only those to destination states whose centre $\vec{C}_\nu=\bra{\nu}\vec{r}\ket{\nu}$ is within a cutoff distance of the current state. The procedure to determine $r_\mathrm{hop}$ is given in steps 1-2 of \cref{listing}, and involves increasing $r_\mathrm{hop}$ successively by one lattice spacing until the sum of all outgoing rates converges to within a desired accuracy $a_\mathrm{dKMC}$. The accuracy of dKMC is therefore tunable by changing $a_\mathrm{dKMC}$ according to the desired precision and available computational resources. This approximation further reduces the number of rates calculated, with only $O(r_\mathrm{hop}^d)$ required at each hop instead of the full $O(N^d)$.

Third, we impose a site contribution cutoff (\cref{fig:approximations_excitons}d). When calculating each Redfield rate, instead of including all site combinations in the damping rates (\cref{eq:Gamma}), we ignore contributions from sites that do not considerably contribute to the populations of the donor and acceptor states. In particular, for each summation index in \cref{eq:Gamma}, we choose the smallest possible number of sites such that the total population of the corresponding exciton state on those sites exceeds a population cutoff (chosen to be $a_\mathrm{dKMC}$); for example, we reduce the sum over $m$ to only go over the smallest possible subset of sites such that  $\sum_m\abs{\braket{\mu|m}}^2>a_\mathrm{dKMC}$.
Doing so ignores contributions from sites where the excitonic state's amplitude is small, significantly reducing the number of site combinations included in each rate calculation. Estimating the precise scaling is difficult, as the spatial extent of each state is unpredictable. However, as an upper bound, the scaling is reduced from $O(N^{4d})$  to no more than $O(r_\mathrm{hop}^{4d})$. 

Fourth, we diagonalise the Hamiltonian on the fly (\cref{fig:approximations_excitons}e). Instead of diagonalising the entire landscape of size $N^{d}$, we only diagonalise a small subsystem containing sites within a radius $r_\mathrm{Ham}$ of the current location of the exciton. $r_\mathrm{Ham}$ is chosen to be as small as possible to reduce computational cost without unduly affecting the diffusion coefficient results, using the procedure in \cref{listing}. This procedure is combined with that for finding $r_\mathrm{hop}$: before each increment of $r_\mathrm{hop}$, we increase $r_\mathrm{Ham}$ until the sum of outgoing rates converges to the desired accuracy $a_\mathrm{dKMC}$. The net result is $r_\mathrm{hop}$ and $r_\mathrm{Ham}$ which reproduce, on average, the fraction $a_\mathrm{dKMC}$ of the total outgoing rate sum. After the new destination state has been selected, a new Hamiltonian subsystem is diagonalised centred at the new location of the exciton. Diagonalising on the fly reduces the cost of finding the polaron states from $O(N^{3d})$ to $O(r_\mathrm{Ham}^{3d}n_\mathrm{hop}n_\mathrm{traj})$. 

Overall, these four approximations reduce the scaling from $O(N^{3d})+O(N^{6d})$ in sPTRE to $O(r_\mathrm{Ham}^{3d}n_\mathrm{hop}n_\mathrm{traj})+O(n_\mathrm{traj}n_\mathrm{hop}r_\mathrm{hop}^{5d})$ in dKMC. For example, for the largest system we study, $\mu=\SI{4}{D}$ in 3D, dKMC reduces the computational cost by 34 orders of magnitude. 

The full dKMC algorithm is included in \cref{listing}. Steps~1-2 involve calibrating the hopping radius $r_\mathrm{hop}$ and the Hamiltonian radius $r_\mathrm{Ham}$. Steps~3-4 then outline the kinetic Monte Carlo procedure to propagate the exciton motion and calculate $D$.

\section{\label{sec:appendix_parameter_choice}Choices of accuracy and end time}

\begin{figure}[b]
    \centering
    \includegraphics[width=\columnwidth]{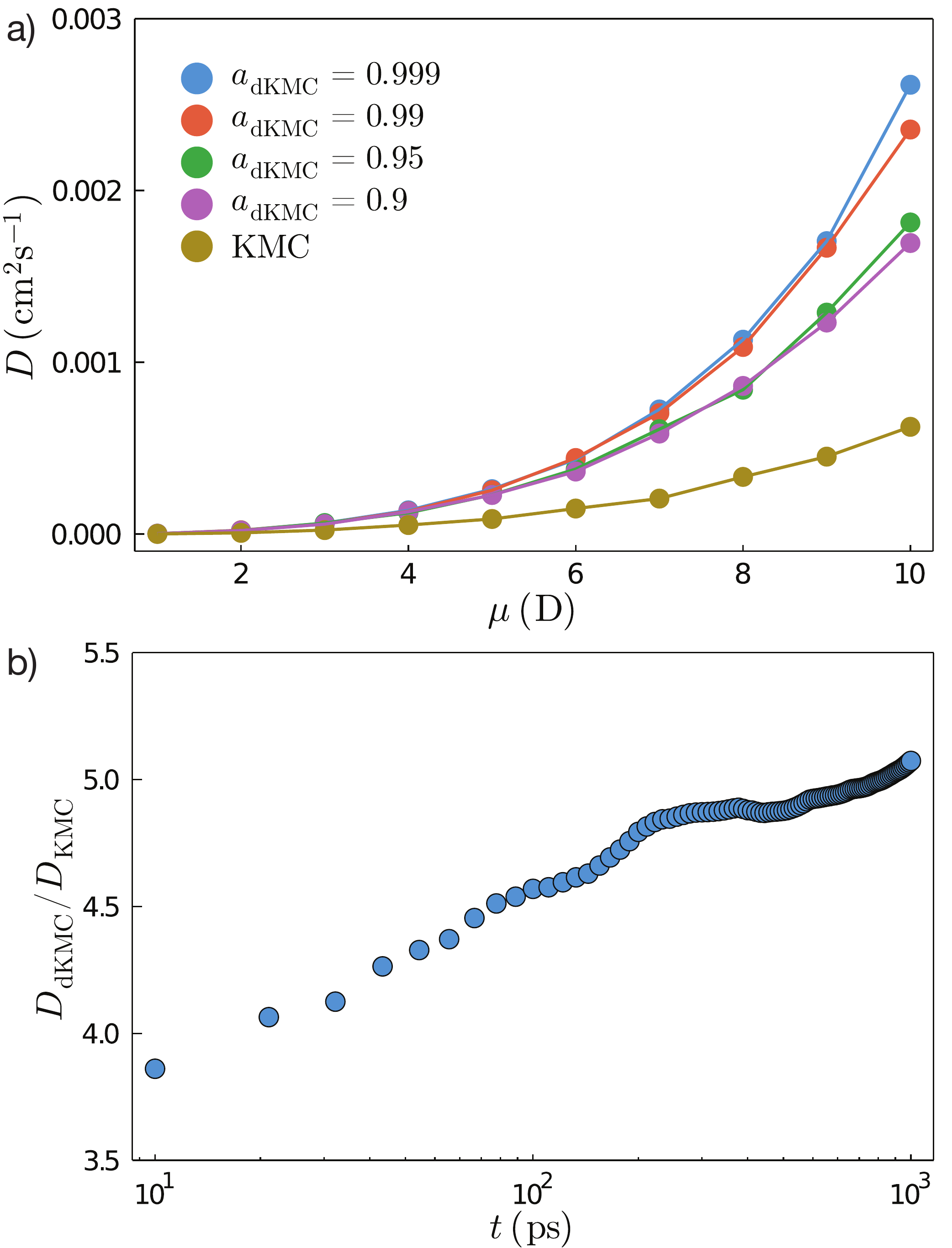}
    \caption{\textbf{Choices of accuracy and end time are conservative.}
    \textbf{a)}~dKMC diffusion coefficient $D$ in one dimension as a function of the transition dipole moment $\mu$ for a range of accuracies $a_\mathrm{dKMC}$. Our choice of $a_\mathrm{dKMC}=0.99$ is conservative because increasing it would only increase $D$ and the delocalisation enhancement of $D$ above the KMC prediction.
    \textbf{b)}~Delocalisation enhancement ($D_\mathrm{dKMC}/D_\mathrm{KMC}$) as a function of time $t$ for $\mu=\SI{5}{D}$ and $\sigma=\SI{30}{meV}$ in 2D. When significant energetic disorder is present, dispersive transport causes diffusion coefficients to be time dependent. We choose a transport time of $t_\mathrm{end}=\SI{100}{ps}$ as a typical timescale in exciton transport, but increasing it would only increase the predicted delocalisation enhancement.
    }
    \label{fig:parameter_choice}
\end{figure}

\Cref{fig:parameter_choice} shows the conservative nature of our choices of two parameters we fixed throughout this work: the accuracy $a_\mathrm{dKMC}$ and the end time $t_\mathrm{end}$. 

\Cref{fig:parameter_choice}a shows the diffusion coefficient $D$ predicted by dKMC as a function of $\mu$ for several accuracies $a_\mathrm{dKMC}$, demonstrating that the accuracy of dKMC can be tuned depending on computational resources and desired accuracy. Throughout this work, for computational efficiency we use $a_\mathrm{dKMC}=0.99$, which reproduces almost all of the result obtained for $a_\mathrm{dKMC}=0.999$ in one dimension. This choice is conservative because it leads to a (small) underestimation of delocalisation effects. 

\Cref{fig:parameter_choice}b shows that the delocalisation enhancement $D_\mathrm{dKMC}/D_\mathrm{KMC}$ increases with time. Diffusion coefficients, and therefore enhancements, are time dependent in disordered materials due to the dispersive nature of the transport. It can take a long time, often longer than the exciton lifetime, for excitons to reach their equilibrium energy $E_\infty=-\sigma^2/k_{\mathrm{B}} T$ and, consequently, for the transport to become truly diffusive. In practice, this means that a transport time $t_\mathrm{end}$ needs to be chosen for diffusion simulations; we choose $t_\mathrm{end}=\SI{100}{ps}$, and \cref{fig:parameter_choice}b shows that extending this cutoff would only further enhance the calculated delocalisation effects.

\end{document}